\documentclass[twocolumn]{aastex61}
\usepackage{amsmath}
\usepackage{graphicx}

\usepackage{epsfig}
\usepackage{natbib}
\usepackage{wasysym}
\usepackage{lipsum}
\usepackage{mathrsfs}
\usepackage{float}
\usepackage{rotating}
\usepackage{graphicx}
\usepackage{epstopdf}
\usepackage{amsmath} 

\newcommand{\angstrom}{\textup{\AA}}
\providecommand{\e}[1]{\ensuremath{\times 10^{#1}}}

\usepackage{color}

\usepackage{gensymb}

\usepackage{filecontents}

\begin{document}

\shorttitle{Habitable Evaporated Planets at the Galactic Center}
\shortauthors{Chen, H., Forbes, J.C., Loeb, A.}

\title{Habitable Evaporated Cores and the Occurrence of Panspermia \\near the Galactic Center}

\author{{\large Howard Chen}}

\affil{{\normalsize Department of Earth \& Planetary Science, Northwestern University, Evanston, IL 60208, USA}}
\affil{{\normalsize Center for Interdisciplinary Exploration \& Research in Astrophysics (CIERA), Evanston, IL 60208, USA}}

\author{{\large John C. Forbes}}

\affil{{\normalsize Institute for Theory and Computation, Harvard University, Cambridge MA 02138, USA}}
\affil{{\normalsize Harvard-Smithsonian Center for Astrophysics, 60 Garden St., Cambridge, MA 02138, USA}}

\author{{\large Abraham Loeb}}
\affil{{\normalsize Institute for Theory and Computation, Harvard University, Cambridge MA 02138, USA}}
\affil{{\normalsize Harvard-Smithsonian Center for Astrophysics, 60 Garden St., Cambridge, MA 02138, USA}}

\correspondingauthor{Howard Chen, Northwestern University}
\email{howard@earth.northwestern.edu, john.forbes@cfa.harvard.edu, aloeb@cfa.harvard.edu}

\begin{abstract}
Black holes growing via the accretion of gas emit radiation that can photoevaporate the atmospheres of nearby planets.
Here we couple planetary structural evolution models of sub-Neptune mass planets to the growth of the Milky way's central supermassive black-hole, Sgr A$^*$ and investigate how planetary evolution is influenced by quasar activity.
We find that, out to ${\sim} 20$ pc from Sgr A$^*$, the XUV flux emitted during its quasar phase can remove several percent of a planet's H/He envelope by mass; in many cases, this removal results in bare rocky cores, many of which situated in the habitable zones (HZs) of G-type stars. Near the Galactic Center, the erosion of sub-Neptune sized planets may be one of the most prevalent channels by which terrestrial super-Earths are created. As such, the planet population demographics may be quite different close to Sgr A$^*$ than in the galactic outskirts. The high stellar densities in this region (about seven orders of magnitude greater than the solar neighborhood) imply that the distance between neighboring rocky worlds is short ($500-5000$~AU). The proximity between potentially habitable terrestrial planets may enable the onset of widespread interstellar panspermia near the nuclei of our galaxy. More generally, we predict these phenomena to be ubiquitous for planets in nuclear star clusters and ultra-compact dwarfs. Globular clusters, on the other hand, are less affected by the central black holes.  
\end{abstract}


\section{Introduction}
\label{sec:intro}

The large population of exoplanets discovered by {\it Kepler} has shown that the architecture and population of planets in the Solar System is unusual--the current sample of known exoplanets is dominated by super-Earths and sub-Neptunes with low masses ($5~M_\oplus \la M_p \la 20~M_\oplus$), low mean densities ($0.3$ g/cm$^3\la \rho_p \la 1.5 $g/cm$^3$), and short-period orbits ($P \la 50$ days) \citep{MullallyEt2015ApJS}.

Due to the exotic environments of the known exoplanets, their volatile-rich atmospheres may be subjected to a host of external physical processes unparalleled among Solar System planets. For instance, scattering and accretion between planets and planetesimals could result in a range of evolutionary outcomes, both from disruption via ``hit-and-run" collisions \citep{Hwang+Chatterjee+etal2017ApJ} and deep heating by planetesimal deposition \citep{ChatterJee+Chen2018ApJ}.

Photoevaporation due to the XUV photons emitted by the host star is another widespread mechanism affecting short-period, low-mass planets.
Models and observations suggest that the primordial characteristics of {\it Kepler} planets has been strongly shaped by XUV/EUV irradiation \citep[e.g.,][]{Owen&Wu2013ApJ,Lopez&Fortney2013ApJ,Owen+Wu2017ApJ}. 
In fact, several groups (e.g., \citealt{VanEylenEt2017,Lehmer+Cating2017ApJ,ZengEt2017arXiv}) recently presented evidence suggesting that the bimodal distribution in the {\it Kepler} population sample \citep{FultonEt2017AJ} is caused by stellar XUV-driven photo-evaporation of their outer envelopes. This indicates that the gaseous-rocky transition at ${\sim} 1.6 R_\oplus$ (e.g., \citealt{Rogers2015ApJ}) is likely a result of atmospheric stripping, as opposed to a secondary planet population born without gaseous envelopes. 
Understanding the prevalence of rocky planets across a single/multiple planet populations is an important step in constraining the composition of exoplanets. More generally, the occurrence of rocky versus gaseous planets has implications for planetary formation, structure, and habitability.

Recently, \citet{Forbes+Loeb2017arXiv} proposed a new mechanism that could potentially sculpt the planet population on yet a larger scale. They argued that the activity of supermassive black holes (SMBH) located at the cores of galaxies can erode the planetary hydrogen-helium atmospheres. This effect is likely most pronounced when the black hole is visible as a quasar, during which the XUV output is maximized. Indeed, ionized accretion flows due to the strong X-ray fluxes have been observed to take place in several distant Active Galactic Nuclei (AGNs) (e.g., \citealt{pounds2003high}).

\citet{Forbes+Loeb2017arXiv} estimate that the effect of XUV radiation from quasars can extend to planets even ${\sim}$kpc away from the centers of galaxies for massive black holes. In the Milky Way there is evidence to suggest that the central supermassive black hole Sgr A$^*$ has recently undergone periods of strong nuclear activity about 6 Myr ago \citep{NicastroEt2016ApJ} and even a mere 110 years ago \citep{ChurazovEt2017arXiv}, and likely experienced even more dramatic events in the past. This possibility provides a compelling motivation to study the evolutionary histories of planets whose observable properties may be influenced by the activity of Sgr A*.

\section{Numerical Setup}
\label{sec:methods}

In this Letter, we employ the state-of-the-science software instrument {\sc mesa} \citep[][version 8845]{PaxtonEt2011ApJS, PaxtonEt2013ApJS} to couple the structural evolution of exoplanets to the XUV flux and spectral evolution of Sgr A$^*$ during its quasar phases.

\subsection{The Planet Model}

Our {\sc mesa} planet model follows the setup described by \citet{Chen+Rogers2016ApJ}. Consistent with several prior planet evolution models (e.g. \citealt{Owen&Wu2013ApJ}), we consider one-dimensional spherically symmetric planets consisting of a heavy-element, silica-iron-magnesium (Si-Fe-Mg) interior surrounded by a hydrogen-helium dominated envelope. The model satisfies the Newtonian requirement for hydrostatic equilibrium:

\begin{equation} 
\frac{dr}{dm} = \frac{1}{4 \pi r^2 \rho}
\end{equation}

\begin{equation}
\frac{dP}{dm} = \frac{G M}{4 \pi r^4}
\end{equation}

\begin{equation}
\frac{d\tau}{dm} = \frac{\kappa}{4 \pi r^2}
\end{equation}

\noindent where $m(r)$ is mass enclosed within radius $r$, $P$ is the pressure, $\tau$ is the optical depth, $\kappa$ is the opacity, and $\rho $ is the density as a function of local pressure and temperature.

For the equations of state suitable for gas giant planetary atmosphere conditions, we adopt the formalism in \citet{SaumonEt1995ApJS}. Unless otherwise stated, we use star-planet metallicities $Z = 0.03$ and helium mass fractions $Y = 0.25$ consistent with Solar values. Low temperature Rosseland opacity tables native to {\sc mesa} are employed for the optical and infrared wavelengths \citep{FreedmanEt2008}. 
We refer the reader to the original {\sc mesa} papers \citep{PaxtonEt2011ApJS,PaxtonEt2013ApJS} for further descriptions of the equations of state and opacity table options. 

The radiative transfer and the treatment of the atmosphere is based on the two-stream approximation of \citet{Guillot&Havel2011A&A} (an option native to {\sc mesa}). In computing the temperature and basic radiative properties of the planet atmosphere, this approximation assumes a grey atmosphere and incident upward and downward fluxes independent of each other. 

All calculations are performed for planets orbiting solar-mass, solar-metallicity stars and we do not model the pre-main sequence phase of the host star. We assume the age of disk clearing relative to the age of the star is 10 Myr. This duration is consistent with the choice of a ${\sim} 10$ Myr cooling timescale prior to turning on mass-loss in our thermal evolution calculations.

The thermal evolution model is coupled to analytical atmospheric escape processes driven by incident XUV photons ($1 \lesssim \lambda \lesssim 1200 \angstrom$). By convention, the energy-limited escape is parameterized as \citep[e.g.,][]{MurrayClayEt2009ApJ,LopezEt2012ApJ}:

\begin{equation} 
\dot{M} = - \frac{\epsilon_{\rm XUV} \pi F_{\rm XUV} R_{\rm XUV}^3}{G M_p K_{\rm tidal}}, \label{elim}
\end{equation}

\noindent where $\epsilon_{\rm XUV} = 0.1$ is the mass-loss efficiency parameter. $F_{\rm XUV}$ is the extreme ultraviolet energy flux impinging on the planet atmosphere. $R_p$ and $M_p$ are planet radius at optical depth $\tau_{\rm visible} = 1$  and the mass of the planet. $R_{\rm XUV}$ is the distance from the center of the planet to the point where the atmosphere is optically thick to XUV photons. 
The coefficient $K_{\rm tidal}\equiv 1-(3R_{\rm XUV})/(2R_{H})+ 1/[2(R_{H}/R_{\rm EUV})^3]$ corrects for tidal forces \citep{ErkaevEt2007A&A}, where $R_H$ is the Hill radius of the planet. 

We calculate $R_{\rm XUV}$ by approximating the difference between $\tau_{\rm visible}=1$ and $\tau_{\rm XUV}=1$ with

\begin{equation}
R_{\rm XUV} \approx R_p + H \ln \left(\frac{P_{\rm photo}}{P_{\rm XUV}}\right)
\end{equation}

\noindent where $H = (k_{\rm B} T_{\rm photo})/(2 m_H g)$ is the atmospheric scale height at the photosphere. $P_{\rm photo}$ and $T_{\rm photo}$ are the pressure and temperature at the visible photosphere.  We analytically determine $P_{\rm XUV}\approx \left(m_HGM_p\right)/\left(\sigma_{\nu_0}R_p^2\right)$ by adopting the photoionization cross-section of hydrogen $\sigma_{\nu_0}=6\times 10^{-18} \left(h\nu_0/13.6~\mathrm{eV}\right)^{-3}~\mathrm{cm^2}$, where we choose a typical XUV energy $h\nu_0=20~\mathrm{eV}$.   

For the stellar high-energy XUV flux impinging on the planet's dayside, we follow \citet{ValenciaEt2010A&A} by adopting the following:

\begin{equation}
F_{\rm XUV} \approx \alpha \left(\frac{t_*}{10^9 {\rm s}}\right)^{-\beta} a^{-2}
\end{equation}

\noindent where $t_*$ is the stellar age in years, $a$ the planet’s orbital distance in AU, and $\alpha$ and $\beta$ are constants according to \citet{RibasEt2005ApJ}. Their best-fit result for G-type stars are $\alpha$ = 29.7 ergs$^{-1}$ cm$^{-2}$ and $\beta = 1.23$.

The planet's core is modeled as the lower boundary of the atmosphere and the core mass-density relations are derived from self-consistent interior models \citep{RogersEt2011ApJ}. The planet also includes time-dependent core heat:

\begin{equation} 
L_{\rm core} = - c_vM_{\rm core}\frac{{\rm d} T_{\rm core}}{{\rm d} t} + L_{\rm radio}.
\label{eqn:Lcore}
\end{equation}

\noindent The first term on the right hand side of Equation~\ref{eqn:Lcore} represents the internal thermal inertia where $c_v = $ 1.0 J K$^{-1}$ g$^{-1}$  is the heat capacity of the core at constant volume (in $\mathrm{ergs\,K^{-1}g^{-1}}$), $M_{\rm core}$ is the mass of the planet's core, and d$T_{\rm core}$/d$t$ is time derivative of the effective (mass-weighted) core temperature. Here we calculate the Lagrangian time derivative of the temperature at the envelope base as representative for this term.  The term $L_{\rm radio}$ represents the contribution of the decay of radio nuclei from $^{232}$Th, $^{238}$U, $^{40}$K, and $^{235}$U. We assume Earth-like abundances and decay rate constants \citep{Hartmann2014}.

\begin{figure}[t] 
\begin{center}
\includegraphics[width=1.\columnwidth]{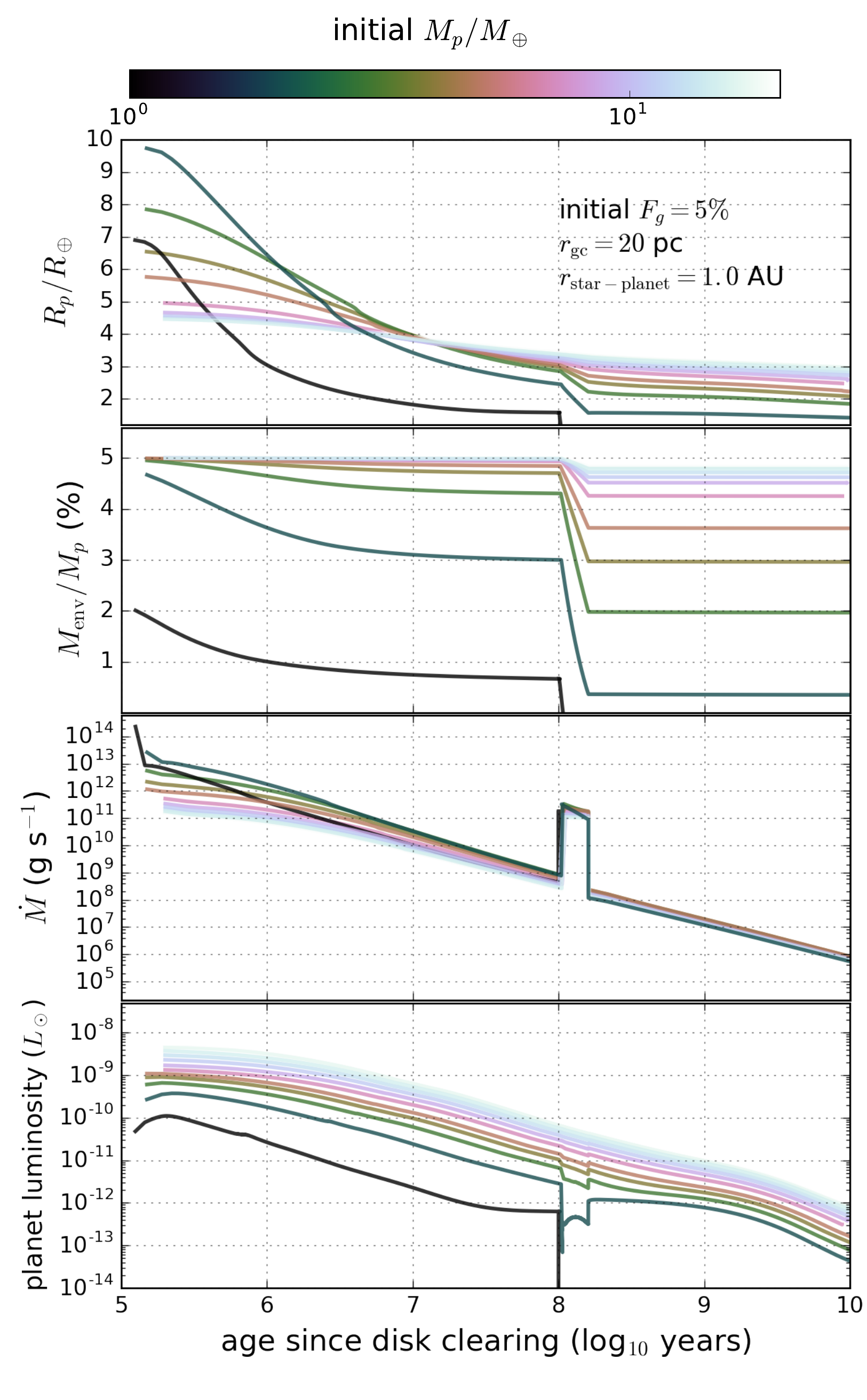}
\caption{\label{fig:pl_evo} Plots of planet radius, gas-envelope mass, mass-loss rate, and photospheric luminosity over time. Simulations are initialized with the same H/He mass fraction $= 5\%$, galactocentric distance of 20~pc, and star-planet separation of 1.0 AU, but with different initial mass as denoted by color. The XUV radiation has a substantial effect on the envelope mass, particularly for low-mass planets. Following the quasar phase of Sgr A*, the remainder of the planetary evolution is evolved via thermal cooling and contraction. Note that for the 1 $M_\oplus$ case, a large fraction of the H/He envelope is lost in the first timestep. }
\end{center}
\end{figure}

\subsection{XUV Flux and Growth of Supermassive Black Holes}

The high-energy radiation incident from Sgr A$^*$ during its quasar episodes can impart significant heating to planetary atmospheres in its vicinity. This flux $F_{\rm BH, XUV}$ can, in some regimes, be much greater than that outputted by the planet's host star $F_{\rm *, XUV}$, depending on the specific star-planet separation and galactocentric distances considered. 

In this study, we couple our {\sc mesa} structural calculations to the predicted \citep{ThomasEt2016ApJ} XUV flux of Sgr A$^*$ in an active phase. The following variables and their respective values were used in deriving the AGN spectra:

\begin{itemize}
\item  corona size (log$_{10}$ in gravitational radii) $r_c =  1.0$
\item  $\eta_{\rm Edd}$ (Eddington ratio)  $=1$
\item  the photon power law index $\Gamma = 1.8$
\item  non-thermal fraction $p_{NT} = 0.1$
\item  the black hole mass $M_{\rm BH} =  4 \times 10^6 M_\odot$ 
\end{itemize}

The choices of $\Gamma, p_NT,$ and $r_g$ are based on the arguments made by section 3.2 of \cite{ThomasEt2016ApJ}. $\eta_{\rm Edd}$ is the direct result of our assumptions about how black holes grow. The SMBH mass is consistent with that of Sgr A* based on observational measurements \citep{BoehleEt2016ApJ}.
We assume that the black hole remains in this quasar state, where $\eta_{\rm Edd}=1$, for a timescale $t_0=50\ \mathrm{Myr}$. This timescale corresponds to an estimate of the typical quasar lifetime \citep[e.g.][]{Wyithe2003ApJ,FanEt2008ApJ}. Although the supermassive black hole is always accreting material at some level, it is consistent with observational constraints to assume that its growth is dominated by the quasar phase. Planets are affected by this radiation only when the XUV flux exceeds $\sim 0.1\ \mathrm{erg}\ \mathrm{s}^{-1}\ \mathrm{cm}^{-2}$ \citep{BolmontEt2017MNRAS}, so the effect on planetary atmospheres will also be dominated by the quasar phase. We have tested a suite of simulations with variable SMBH luminosity and found that the inclusion of variability does not alter the calculated planet observables by more then 5\%, which is much lower than current observational uncertainties. We do not account for the attenuation of the radiation passing through the AGN torus and the interstellar medium.

\begin{figure}[t] 
\begin{center}
\includegraphics[width=1.\columnwidth]{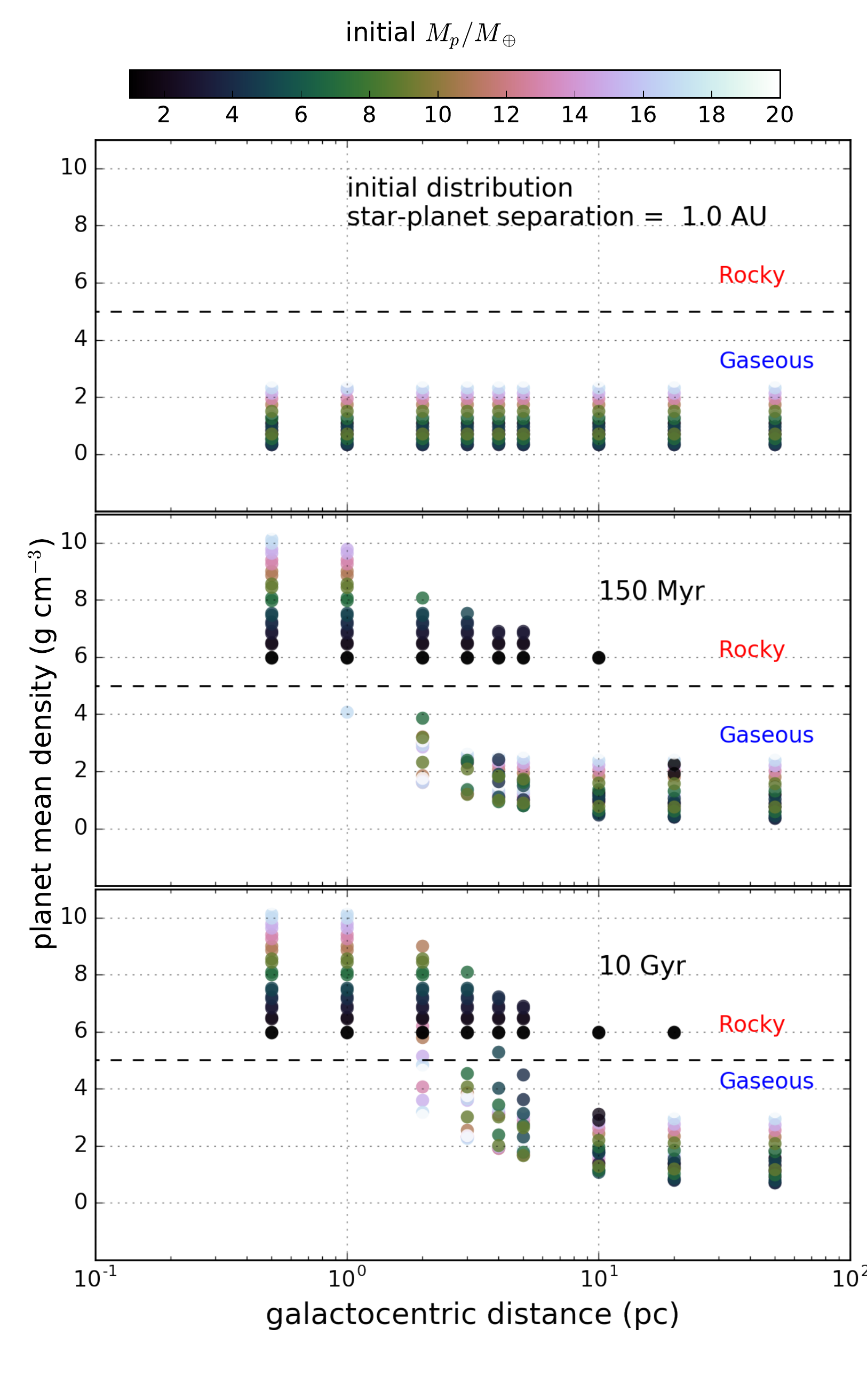}
\caption{\label{fig:pl_pop} Planet mean density $\rho_p$ as a function of galactocentric distance $r_{\rm gc}$. Accompanying the decrease of $R_{\rm gc}$, there is a substantial rise in the number of HECs. Conversely, it becomes increasingly difficult to completely evaporate sub-Neptunes for orbital distances beyond 20~pc from Sgr A$^*$.}
\end{center}
\end{figure}

\section{Results}
\label{sec:results}

The coupled thermal/mass-loss evolution simulations, for fixed H/He mass fraction $F_g =  5\%$ and star-planet separation = 1.0 AU, are shown in Figure~\ref{fig:pl_evo}. Planet radius, envelope mass fraction, mass-loss rate, and luminosity as a function of planet age (from $10^5$ to $10^{10}$ years since disc clearing) are shown in four separate panels. The AGN flux begins at a planet age of 100~Myr and lasts for 50~Myr. 

From our calculations, the XUV flux from the host star (${\sim} 5.6\  \mathrm{erg}\ \mathrm{s}^{-1}\ \mathrm{cm}^{-2}$ at 1.0 AU, and ${\sim} 850\  \mathrm{erg}\ \mathrm{s}^{-1}\ \mathrm{cm}^{-2}$ at 0.1 AU ) is typically much lower than that from the SMBH (${\sim} 4\e{4}\  \mathrm{erg}\ \mathrm{s}^{-1}\ \mathrm{cm}^{-2}$ at 10 pc). The periods of high stellar XUV approaching that of the SMBH only persist for a short (${\sim} 1000$ year) time. This is consistent with expectations at these small galactocentric distances $R_{\rm gc} \la 100$ pc. Outside of the Sgr A$^*$ quasar phases, planetary radii decrease over time due to atmospheric escape and thermal contraction. Mass is initially lost at large rates due to both the higher stellar XUV output and planet entropy (and thus cross-sectional radii) at young ages. The photospheric luminosity of the planet falls gradually as the planet cools. During the quasar phase, the mass-loss rate increases by orders of magnitude, and the photospheric luminosity is also perturbed by the abrupt increase in mass-loss rate, which induces sudden changes in the XUV-photon radii of the planet models. The large effect of the quasar as compared to the host star is expected, since planets at 1.0 AU receive much lower XUV irradiation than the typical planet observed by {\it Kepler}.

We have established that XUV irradiation from Sgr A$^*$ can alter the evolution pathways of low-mass planets, but it is unknown to what extent this may affect the composite planet population. To examine this, we run a grid of planet mass $M_p$ from 1 to 25 $M_\oplus$ and H/He mass fractions between $F_g =  5\%$ and 10\%. These choices reflect the most common planet masses from {\it Kepler} statistics. For instance, \citet{HowardEt2010Science} showed  that  the  planet
occurrence  rate  increased  towards  low-masses (peaking at ${\sim}3-15 M_\oplus$). These choices also reflect the most probable initial H/He inventory from planet formation models (e.g., \citealt{Bodenheimer&Lissauer2014ApJ}).
With each combination of $M_p$ and $F_g$, we explore 11 galactocentric distances $R_{\rm gc} = 0.1,\ 1,\ 2,\ 3,\ 4,\ 5,\ 10,\ 15,\ 20,\ 50,\ \mathrm{and}\ 100$ pc.

These results are shown in Figure~\ref{fig:pl_pop}, where we illustrate how the initial planet distribution is sculpted by the AGN XUV flux at star-planet separation of 1.0 AU. Panel a shows our initial planet distribution setup unaltered by any post-formation processes. Panel b illustrates the case at 150 Myr planet age, at which point the AGN has just turned off. Panel c is the final planet distribution at 10~Gyr planet age. One can see that most of the shifts in planet composition arise from the activity of Sgr A$^*$; the set of models at the largest galactocentric distance are largely unchanged from the initial conditions, while planets close to the quasar are completely transformed from gaseous to rocky\footnote{We defined a planet being ``rocky" when $F_g\lesssim 10^{-5}$. At this point, we terminate the code and assume the remainder of the H/He is quickly lost via Jeans escape.}.  

Due to the evolving stellar $F_{\rm XUV}$ and planet radius, the onset of the AGN at various times may have substantial impacts on the outcomes of planet density. To gauge the relative importance of this effect, we ran simulations with fixed initial $M_p = 5~M_\oplus$ and $F_g = 5\%$. We vary the galactocentric distance as in the previous set of simulations, but now we vary the AGN occurrence times between 10~Myr and 1~Gyr. The aim is to allow for uncertainty in the actual time of the AGN and a range of star/planet formation times. We also explore three star-planet separation $a = 0.7$, 1.0, and 1.5 AU, bracketing the optimistic habitable zone for Sun-like stars \citep{KastingEt1993Icarus,YangEt2014ApJL}.

\begin{figure}[t] 
\begin{center}
\includegraphics[width=1\columnwidth]{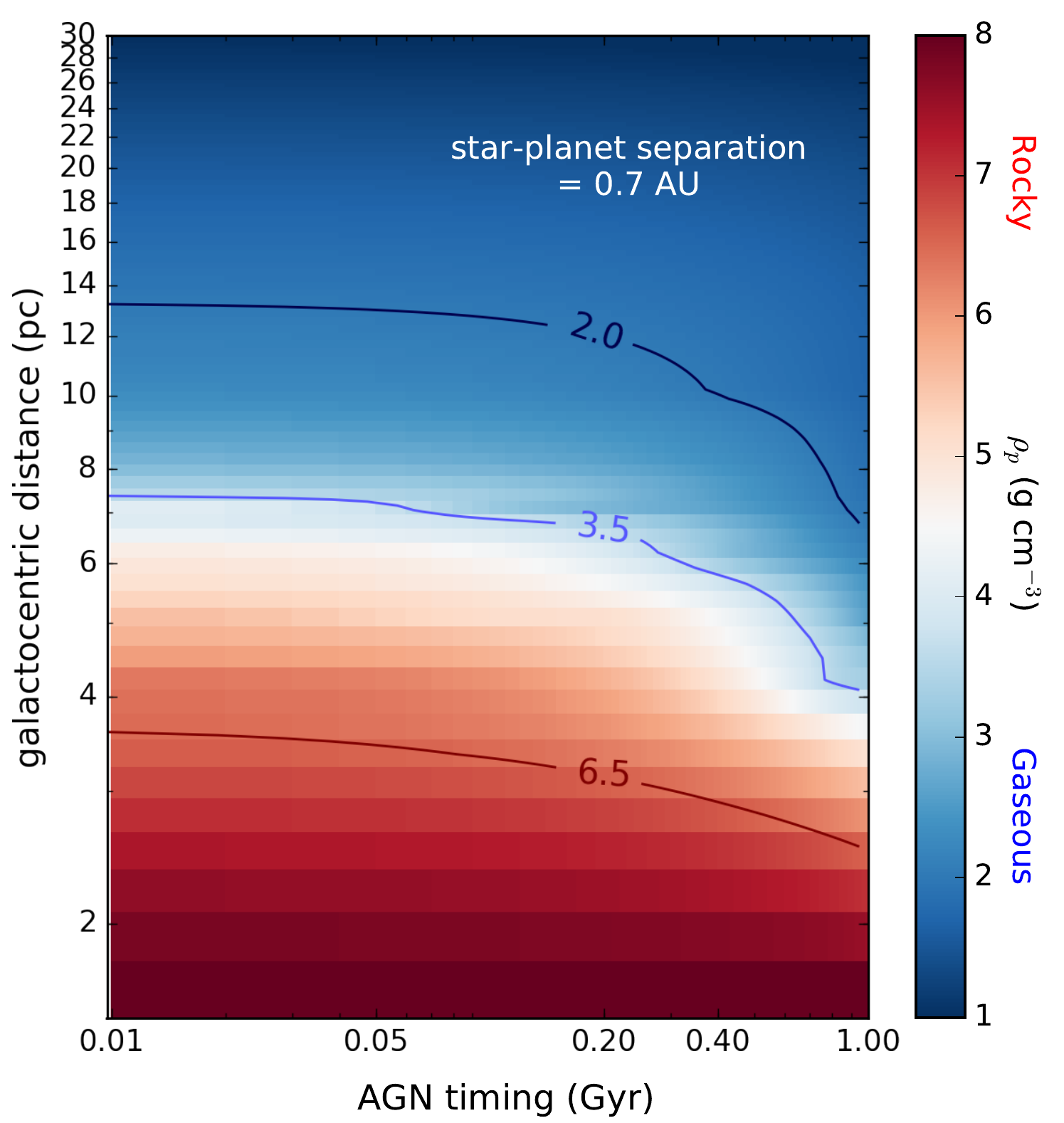}
\caption{\label{fig:pl_pop2} Color-coded contour plot illustrating maximum planet density achieved as a function of AGN occurrence times and galactocentric distance, for $M_p = 5-10 M_\oplus$ and $F_g = 1-5\%$. The highest density region (lower left-hand corner) reflects the most efficient parameter space regime for HEC-transformation. Not surprisingly, this regime correspond to the the corner of the parameter space where both the x and y-axis variables are minimized. This means that the probability of a given planet model being rocky increases with decreasing galactocentric radius and greater planet age. At the inner edge of the habitable zone (0.7~AU case shown here), more areas are converted to rocky compared to the outer edge of the habitable zone due to the increased XUV irradiation from the host star.}
\end{center}
\end{figure}

Figure~\ref{fig:pl_pop2} illustrate the effects of different AGN occurrence times on the final planet composition. We have already seen that the final planet composition is strongly dependent on galactocentric distance; we show here that it is also weakly dependent on the AGN occurrence time and star-planet separation. AGN activity early in the planet's evolution (${\sim}$ Myr) has substantially different resultant planet density imprinted on the population than turning it on at a much later time (${\sim}$ Gyr). The highest density region of Figure~\ref{fig:pl_pop2} reflects the most efficient parameter space regime for HEC-transformation. 

\section{Discussion}
\label{sec:disc}

Prior to discussing the implications for various aspects of planetary science, the reader should bear in mind a few caveats that may affect our numerical calculations. First, to limit the potentially huge parameter space, we only adopted the XUV evolution model of G-type stars \citep{RibasEt2005ApJ}. However, lower mass stars such as M-dwarfs have much longer XUV luminosity saturation timescales (see e.g., \citealt{Chabrier+Baraffe1997}) compared to the G-dwarf models employed here. 
Therefore, the number of predicted rocky planets should increase with altered stellar spectra of M-dwarfs. On a similar note, the orbital semi-major axis of all planets simulated have $\ga 0.7$ AU, which lie in a regime that is distinctively not {\it Kepler}-like. At closer-in orbital separations of $\la 0.1 $ AU, we predict that the planet population would be more dominated by evaporated (but non-habitable) cores.

Second, we restrict our {\sc mesa} simulations to planets with H/He-rich atmospheres and do not consider, for example, diffusion-limited escape processes that may be important for atmospheric structures with much less H/He \citep[e.g.,][]{KastingEt2015ApJL} (though we do apply the energy-limited scheme to terrestrial atmospheres, see Section~4.2). 
Third, we ignore all other effects that AGNs may have on atmospheric chemistry. Black hole XUV irradiation would not only drive enhanced hydrodynamic escape, which we have solely considered here. We will determine the effects of X-ray irradiation on photochemistry of planetary atmospheres in a separate paper using 3-D global chemistry models.

\subsection{Implications for Habitability}

Figure~\ref{fig:pl_pop} indicates that, for $R_{\rm gc} \la 20$ pc, planets with up to 10\% of H/He by mass can be stripped to bare cores in the HZs of solar-type stars, resembling ``habitable evaporated cores" (HECs) found by the models of \cite{LugerEt2015AsBio}\footnote{\citet{LugerEt2015AsBio} found that a subset of sub-Neptune mass planets initially accreted some amount of H/He can be evaporated down to bare cores. The phrase ``habitable evaporated cores" (HECs) was used to describe these remnant cores if they lie in the circumstellar habitable zones of their host M-dwarf stars \citep{KastingEt1993Icarus}. In referring to them as ``habitable", we are of course ignoring the various other atmospheric, chemical, climatic, and geophysical affects that may play critical roles in governing habitability. We are following the naming convention of \cite{LugerEt2015AsBio}, to distinguish these planets from true terrestrial planets formed by the aggregate of planetesimals in the HZs.}, as rocky planets are more likely vessels to harbor life as we know it. At $R_{\rm gc} \ga 30$~pc, only planets with 1-2 $M_\oplus$ can be evaporated completely at star-planet separations of 1.0 AU. This is consistent with the findings of \citet{LammerEt2014MNRAS} that planets more massive than ${\sim} 1.5 M_\oplus$ typically cannot lose their accreted gas. Even for planets around M-dwarfs, it becomes increasingly difficult to evaporate planets with rocky cores significantly beyond $1.5~M_\oplus$ \citep{LugerEt2015AsBio}. Our simulations show that at $R_{\rm gc} \la 10$ pc, planets larger than ${\sim} 5 M_\oplus$ can still be completely eroded. It is also possible that planets simply are not born with much H/He due to photoevaporation of the proptoplanetary disc by the black hole \citep{Murray-Clay+Loeb2012}. Yet, the timescale window to allow this to happen is arguably much smaller than the one studied here-- in Figure~\ref{fig:pl_pop2} we demonstrated that complete erosion of H/He is possible throughout the planets' lifetime.

Many authors (e.g., \citealt{LineweaverEt2004Science}) have questioned the habitability of planetary systems so close to the Galactic Center. These authors argue that planets here are especially vulnerable to supernova explosions and gamma-ray bursts. Although these events may be harmful to the biosphere via their effect on atmospheric chemistry such as O$_3$, the distance out to which they have strongly deleterious effects on habitability is controversial. Another argument against habitability close to the Galactic Center is the dynamical influence of the high stellar density such as the passing of neighboring stars. Indeed, calculations indicate that planets beyond ${\sim}0.5-10$ AU from the host star are on extremely unstable orbits \citep{Sigurdsson1992ApJ}, which would deter habitable conditions around the HZs of G and K-type stars. For M-dwarfs however, their HZs are so close (0.03-0.1 AU) such that the attending planets may be insulated from catastrophic disruption and be relatively safe from stellar interactions. A similar problem is that terrestrial planets must survive dynamical ejection from gas giant planets \citep{BarclayEt2017ApJ}.
It could be very well be true that terrestrial planets directly assembled from planetary embryo are susceptible to ejection early on. As argued in this paper, new rocky planets may appear later in the life of the system via the transformation of mini-Neptunes to rocky cores by the AGN.

Habitability of HECs and initially-terrestrial planets may also be threatened by further XUV radiation by the central SMBH. 
With a simplified energy-limited escape equation, we find that terrestrial atmospheres may lose up to $5 \times 10^{23}$ g (${\sim}100~M_{\rm atm, \oplus}$) worth of atmospheric mass within 20 pc of Sgr A*. In light of these numbers, we may expect the atmospheres of terrestrial planets in the vicinity of SMBHs to be significantly, if not entirely, eroded. However, there are several caveats. First, the energy-limited escape scheme was initially derived to estimate the escape of H/He. The extent to which this equation can be applied to nitrogen-dominated atmospheres is uncertain. More specifically, the heating efficiency parameter $\epsilon$ may be low for atmospheres with higher mean-molecular weight (e.g., \citealt{Owen&Jackson2012ApJ,Owen&Wu2013ApJ}). Therefore, applying this equation to estimate mass-loss rates from Earth-like planets may result in in severe over-predictions. 
Moreover, even if terrestrial atmospheres are eroded by the XUV flux of the SMBH, secondary atmospheres may arise later from volcanism or the impacts of volatile-rich remnant planetesimals after the quasar phase has passed.


\begin{figure}[t] 
\begin{center}
\includegraphics[width=1.1\columnwidth]{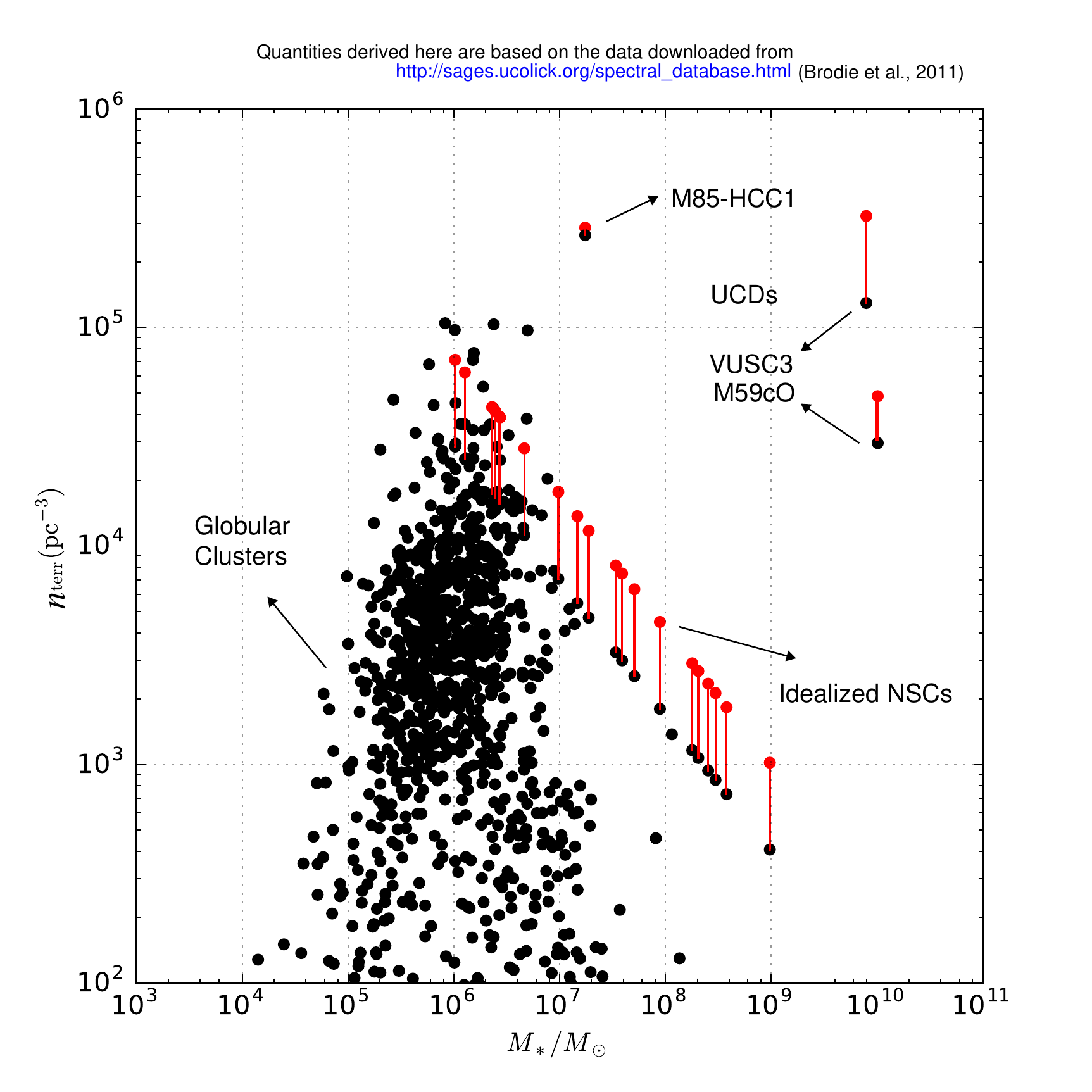}
\caption{\label{fig:panspermia} Theoretical spatial density of rocky planets ($n_{\rm terr}$) as a function of stellar mass (in $M_\odot$) of the system.  We use the visual magnitudes and effective radii of a collection of nearby spheroidal stellar systems compiled by \citet{Brodie2011AJ} to estimate the stellar densities, stellar masses, and black hole masses for each system. We assume a mass to light ratio of 2, and a black hole mass following \citet{Haering+Rix2004}. We also estimate the mass and density of nuclear star clusters following the simplified model of \citet{MacLeodEt2012ApJ}. The most significant effect due to SMBH-induced erosion are planets situated within ultra-compact dwarfs (UCDs) and nuclear star clusters (NSCs). Interestingly, systems that are shifted most substantially also correspond to systems with the largest estimated mass ($M_*/M_\odot$). Other populations are not significantly affected as they are more dispersed or lie farther from their respective central black holes. }
\end{center}
\end{figure}

\subsection{Implications for Panspermia}

The concept of panspermia, where life is transmitted via interplanetary or interstellar contact, has been discussed for decades (e.g., \citealt{Chyba+Sagan1992,Lin+Loeb2015ApJL}). Here we shall comment on what our results imply for such processes. Several authors (e.g., \citealt{Lin+Loeb2015ApJL}) argued that stellar density is indirectly linked to panspermia efficiency. This is because high density implies higher likelihood of close encounters and thus transfer events between planets, as in globular clusters \citep{Stefano+Ray2016ApJ} or even within the same system \citep[e.g. TRAPPIST-1][]{Lingam+Loeb2017PNAS}.

Recent observations by the VLT (Very Large Telescope) argue for a high stellar density close to Sgr A$^*$. Within 5~pc of the galactic center, the stellar mass density is estimated to be ${\sim} 10^5 M_\odot {\rm pc}^{-3}$ \citep{SchodelEt2017arXiv}, with a power-law index of $\gamma \approx 1.23$. (By contrast, the spatial stellar density in the solar neighborhood is 0.2 $M_\odot {\rm pc}^{-3}$.) Within 5~pc of the galactic center, the mean distance between neighboring stars would be a mere ${\sim} 5000$~AU (compared to the closest Oort cloud objects at ${\sim} 6000$~AU from the Earth). Oort cloud comets can be perturbed onto Earth-crossing trajectories \citep{Fernandez+Ip1987}, highlighting the relative ease for cometary or asteroidal exchanges between planetary systems at these distances. In addition, if the processes of panspermia occur outsode of active periods, then the chances for microbes to survive the interstellar journey are greatly enhanced. 

We now turn to exploring how the AGN effects considered here may impact the broader planet distribution across a variety of stellar systems. Figure~\ref{fig:panspermia} shows the results for the predicted spatial number density of terrestrial planets $n_{\rm terr}$ against the typical mass of each stellar system. To analyze the degree of AGN effects, we show our results as both an initial estimate (black dot) and a new estimate taking into account the transformation of gaseous planets to rocky as we have described in this work (red dot). We assume unity for the planet to star number density ratio and that the initial rocky to gaseous planet ratio is 2 to 3 (40\% rocky vs 60\% gaseous). 
Among all objects shown, globular clusters, ultra-compact dwarfs (UCDs), and low-mass NSCs all reach of order $\sim 10^5$ rocky planets per pc$^{3}$. The density of rocky planets is not substantially increased in globular clusters by the putative black holes, but UCDs and NSCs are each close enough to a sufficiently massive black hole for their mini-Neptunes to be affected.


\subsection{Implications for Observations}


Observing the effects we have discussed here will require detecting and characterizing planets formed in the Galactic Center. This region is a challenging environment for detecting planets owing to the distance, crowding, extinction, and foregrounds. The radial velocity method is generally limited to nearby bright stars to achieve high signal-to-noise in high-resolution spectra. Even though gravitational microlensing can detect down to Mars-mass objects up to ~8 kpc away \citep{Gould+Loab1993ApJ,BarclayEt2017ApJ}, crowding and extinction compel upcoming microlensing surveys to avoid the central degree of the Galaxy \citep{SpergelEt2015arXiv}. Rather than detecting planets in the galactic center, it may also be possible to identify nearby stars that have migrated through the galactic disk \citep[see e.g.][]{LoebmanEt2017ApJL} via their chemical abundances, e.g. elevated [$\alpha$/Fe], and characterize the planets around such stars. However, the stars that have migrated from the central ~20 pc of the Galaxy will likely be rare and difficult to distinguish from stars that have migrated from slightly larger galactocentric distances. We therefore expect that the best prospect for detection is with the next generation of ground-based telescopes.


We can estimate the apparent magnitude $m$ of a target star assuming, for a Sun-like star, an absolute magnitude $M = 4.8$, distance $d = 8$ kpc, and extinction $A(\lambda) \approx 3.3$ in the IR. With these values, we find $m \approx 17.6$, whereas the limiting magnitude on the E-ELT's IR camera (MICADO) after 5 hours of integration time in the J, H, and K bands are 30.8, 30.8 and 29.8 respectively \citep{DavisEt2010SPIE}. Additionally, the photometric uncertainties calculated by \citet{GullieuszikEt2014A&A} in the reddest bands of I and H are lower than ${\sim} 1\e{-3}$. This value is lower than the $(R_p/R_*)^2 \sim 2\e{-3}- 2.5\e{-3}$ for a sub-Neptune sized HEC transiting a Sun-like star. These numbers indicate that assessing our predictions is within the realm of possibility using an ELT-like telescope. Ultimately however, it will depend on whether large telescopes (${\sim} 30$ m) have sufficient angular resolution to resolve individual stars in this crowded environment.

Finally, a more speculative test is to search for a saturation of extraterrestrial colonies $\la 1$ kpc from the Galactic Center. We suggest searching for signals between 1300 and 1800 MHz $\la 1$ kpc from Sgr A* based on the arguments of versed literature (e.g., \citealt{Cocconi+MorrisonNATURE}).



\section{Conclusion}

In this Letter, we have presented a comprehensive study of coupled AGN growth models to the thermo-physical simulations of sub-Neptune sized planets. We have followed the arguments made by \citet{Forbes+Loeb2017arXiv} and used a theoretical spectra of Sgr A$^*$ to determined the amount of ionizing flux incident upon a range of {\sc mesa} planet structural models. Our models predict that the XUV flux from Sgr A$^*$ during quasar phase can blow off several percent of the H/He envelopes of mini-Neptunes in the HZs of Sun-like stars and, in several cases, lead to completely barren cores (``HECs", \citealt{LugerEt2015AsBio}). This prediction arises naturally from the consequences of black hole accretion and planet evaporation theories. In addition, it can be tested by identifying migrated systems with high-[$\alpha$/H], searching for ET signals in the microwave $\la 1$ kpc from Sgr A*, and detecting transiting HECs with the E-ELT in the I and H bands.


\acknowledgements
We thank the anonymous referee for insightful feedback that substantially improved the quality of the manuscript. We thank Manasvi Lingam for helpful comments regarding planetary habitability. We also thank Harvard's Black Hole Initiative which is funded by a grant from the John Templeton Foundation. JCF and AL acknowledge the Breakthrough Prize Foundation for support.

\software{MESA (v8845; Paxton et al. 2011, 2013)}


\begin{thebibliography}{}
\expandafter\ifx\csname natexlab\endcsname\relax\def\natexlab#1{#1}\fi

\bibitem[{{Barclay} {et~al.}(2017){Barclay}, {Quintana}, {Raymond}, \&
  {Penny}}]{BarclayEt2017ApJ}
{Barclay}, T., {Quintana}, E.~V., {Raymond}, S.~N., \& {Penny}, M.~T. 2017,
  \apj, 841, 86

\bibitem[{{Bodenheimer} \& {Lissauer}(2014)}]{Bodenheimer&Lissauer2014ApJ}
{Bodenheimer}, P., \& {Lissauer}, J.~J. 2014, \apj, 791, 103

\bibitem[{{Boehle} {et~al.}(2016){Boehle}, {Ghez}, {Sch{\"o}del}, {Meyer},
  {Yelda}, {Albers}, {Martinez}, {Becklin}, {Do}, {Lu}, {Matthews}, {Morris},
  {Sitarski}, \& {Witzel}}]{BoehleEt2016ApJ}
{Boehle}, A., {Ghez}, A.~M., {Sch{\"o}del}, R., {et~al.} 2016, \apj, 830, 17

\bibitem[{{Bolmont} {et~al.}(2017){Bolmont}, {Selsis}, {Owen}, {Ribas},
  {Raymond}, {Leconte}, \& {Gillon}}]{BolmontEt2017MNRAS}
{Bolmont}, E., {Selsis}, F., {Owen}, J.~E., {et~al.} 2017, \mnras, 464, 3728

\bibitem[{{Brodie} {et~al.}(2011){Brodie}, {Romanowsky}, {Strader}, \&
  {Forbes}}]{Brodie2011AJ}
{Brodie}, J.~P., {Romanowsky}, A.~J., {Strader}, J., \& {Forbes}, D.~A. 2011,
  \aj, 142, 199

\bibitem[{{Chabrier} \& {Baraffe}(1997)}]{Chabrier+Baraffe1997}
{Chabrier}, G., \& {Baraffe}, I. 1997, \aap, 327, 1039

\bibitem[{{Chatterjee} \& {Chen}(2018)}]{ChatterJee+Chen2018ApJ}
{Chatterjee}, S., \& {Chen}, H. 2018, \apj, 852, 58

\bibitem[{{Chen} \& {Rogers}(2016)}]{Chen+Rogers2016ApJ}
{Chen}, H., \& {Rogers}, L.~A. 2016, \apj, 831, 180

\bibitem[{{Churazov} {et~al.}(2017){Churazov}, {Khabibullin}, {Sunyaev}, \&
  {Ponti}}]{ChurazovEt2017arXiv}
{Churazov}, E., {Khabibullin}, I., {Sunyaev}, R., \& {Ponti}, G. 2017, \mnras,
  471, 3293

\bibitem[{{Chyba} \& {Sagan}(1992)}]{Chyba+Sagan1992}
{Chyba}, C., \& {Sagan}, C. 1992, \nat, 355, 125

\bibitem[{{Cocconi} \& {Morrison}(1959)}]{Cocconi+MorrisonNATURE}
{Cocconi}, G., \& {Morrison}, P. 1959, \nat, 184, 844

\bibitem[{{Davies} {et~al.}(2010){Davies}, {Ageorges}, {Barl}, {Bedin},
  {Bender}, {Bernardi}, {Chapron}, {Clenet}, {Deep}, {Deul}, {Drost},
  {Eisenhauer}, {Falomo}, {Fiorentino}, {F{\"o}rster Schreiber}, {Gendron},
  {Genzel}, {Gratadour}, {Greggio}, {Grupp}, {Held}, {Herbst}, {Hess},
  {Hubert}, {Jahnke}, {Kuijken}, {Lutz}, {Magrin}, {Muschielok}, {Navarro},
  {Noyola}, {Paumard}, {Piotto}, {Ragazzoni}, {Renzini}, {Rousset}, {Rix},
  {Saglia}, {Tacconi}, {Thiel}, {Tolstoy}, {Trippe}, {Tromp}, {Valentijn},
  {Verdoes Kleijn}, \& {Wegner}}]{DavisEt2010SPIE}
{Davies}, R., {Ageorges}, N., {Barl}, L., {et~al.} 2010, in \procspie, Vol.
  7735, Ground-based and Airborne Instrumentation for Astronomy III, 77352A

\bibitem[{{Di Stefano} \& {Ray}(2016)}]{Stefano+Ray2016ApJ}
{Di Stefano}, R., \& {Ray}, A. 2016, \apj, 827, 54

\bibitem[{{Erkaev} {et~al.}(2007){Erkaev}, {Kulikov}, {Lammer}, {Selsis},
  {Langmayr}, {Jaritz}, \& {Biernat}}]{ErkaevEt2007A&A}
{Erkaev}, N.~V., {Kulikov}, Y.~N., {Lammer}, H., {et~al.} 2007, \aap, 472, 329

\bibitem[{{Fan} {et~al.}(2008){Fan}, {Lapi}, {De Zotti}, \&
  {Danese}}]{FanEt2008ApJ}
{Fan}, L., {Lapi}, A., {De Zotti}, G., \& {Danese}, L. 2008, \apjl, 689, L101

\bibitem[{{Fernandez} \& {Ip}(1987)}]{Fernandez+Ip1987}
{Fernandez}, J.~A., \& {Ip}, W.-H. 1987, \icarus, 71, 46

\bibitem[{{Forbes} \& {Loeb}(2017)}]{Forbes+Loeb2017arXiv}
{Forbes}, J.~C., \& {Loeb}, A. 2017, ArXiv e-prints, arXiv:1705.06741

\bibitem[{{Freedman} {et~al.}(2008){Freedman}, {Marley}, \&
  {Lodders}}]{FreedmanEt2008}
{Freedman}, R.~S., {Marley}, M.~S., \& {Lodders}, K. 2008, \apjs, 174, 504

\bibitem[{{Fulton} {et~al.}(2017){Fulton}, {Petigura}, {Howard}, {Isaacson},
  {Marcy}, {Cargile}, {Hebb}, {Weiss}, {Johnson}, {Morton}, {Sinukoff},
  {Crossfield}, \& {Hirsch}}]{FultonEt2017AJ}
{Fulton}, B.~J., {Petigura}, E.~A., {Howard}, A.~W., {et~al.} 2017, \aj, 154,
  109

\bibitem[{{Gould} \& {Loeb}(1992)}]{Gould+Loab1993ApJ}
{Gould}, A., \& {Loeb}, A. 1992, \apj, 396, 104

\bibitem[{{Guillot} \& {Havel}(2011)}]{Guillot&Havel2011A&A}
{Guillot}, T., \& {Havel}, M. 2011, \aap, 527, A20

\bibitem[{{Gullieuszik} {et~al.}(2014){Gullieuszik}, {Greggio}, {Falomo},
  {Schreiber}, \& {Uslenghi}}]{GullieuszikEt2014A&A}
{Gullieuszik}, M., {Greggio}, L., {Falomo}, R., {Schreiber}, L., \& {Uslenghi},
  M. 2014, \aap, 568, A89

\bibitem[{{H{\"a}ring} \& {Rix}(2004)}]{Haering+Rix2004}
{H{\"a}ring}, N., \& {Rix}, H.-W. 2004, \apjl, 604, L89

\bibitem[{Hartmann(2015)}]{Hartmann2014}
Hartmann, W. 2015, J. Geophys. Res. Atmos., 120, 5775

\bibitem[{{Howard} {et~al.}(2010){Howard}, {Marcy}, {Johnson}, {Fischer},
  {Wright}, {Isaacson}, {Valenti}, {Anderson}, {Lin}, \&
  {Ida}}]{HowardEt2010Science}
{Howard}, A.~W., {Marcy}, G.~W., {Johnson}, J.~A., {et~al.} 2010, Science, 330,
  653

\bibitem[{{Hwang} {et~al.}(2018){Hwang}, {Chatterjee}, {Lombardi}, {Steffen},
  \& {Rasio}}]{Hwang+Chatterjee+etal2017ApJ}
{Hwang}, J., {Chatterjee}, S., {Lombardi}, Jr., J., {Steffen}, J.~H., \&
  {Rasio}, F. 2018, \apj, 852, 41

\bibitem[{{Kasting} {et~al.}(2015){Kasting}, {Chen}, \&
  {Kopparapu}}]{KastingEt2015ApJL}
{Kasting}, J.~F., {Chen}, H., \& {Kopparapu}, R.~K. 2015, \apjl, 813, L3

\bibitem[{{Kasting} {et~al.}(1993){Kasting}, {Whitmire}, \&
  {Reynolds}}]{KastingEt1993Icarus}
{Kasting}, J.~F., {Whitmire}, D.~P., \& {Reynolds}, R.~T. 1993, Icarus, 101,
  108

\bibitem[{{Lammer} {et~al.}(2014){Lammer}, {St{\"o}kl}, {Erkaev}, {Dorfi},
  {Odert}, {G{\"u}del}, {Kulikov}, {Kislyakova}, \&
  {Leitzinger}}]{LammerEt2014MNRAS}
{Lammer}, H., {St{\"o}kl}, A., {Erkaev}, N.~V., {et~al.} 2014, \mnras, 439,
  3225

\bibitem[{{Lehmer} \& {Catling}(2017)}]{Lehmer+Cating2017ApJ}
{Lehmer}, O.~R., \& {Catling}, D.~C. 2017, \apj, 845, 130

\bibitem[{{Lin} \& {Loeb}(2015)}]{Lin+Loeb2015ApJL}
{Lin}, H.~W., \& {Loeb}, A. 2015, \apjl, 810, L3

\bibitem[{{Lineweaver} {et~al.}(2004){Lineweaver}, {Fenner}, \&
  {Gibson}}]{LineweaverEt2004Science}
{Lineweaver}, C.~H., {Fenner}, Y., \& {Gibson}, B.~K. 2004, Science, 303, 59

\bibitem[{{Lingam} \& {Loeb}(2017)}]{Lingam+Loeb2017PNAS}
{Lingam}, M., \& {Loeb}, A. 2017, Proceedings of the National Academy of
  Science, 114, 6689

\bibitem[{{Loebman} {et~al.}(2016){Loebman}, {Debattista}, {Nidever}, {Hayden},
  {Holtzman}, {Clarke}, {Ro{\v s}kar}, \& {Valluri}}]{LoebmanEt2017ApJL}
{Loebman}, S.~R., {Debattista}, V.~P., {Nidever}, D.~L., {et~al.} 2016, \apjl,
  818, L6

\bibitem[{{Lopez} \& {Fortney}(2013)}]{Lopez&Fortney2013ApJ}
{Lopez}, E.~D., \& {Fortney}, J.~J. 2013, \apj, 776, 2

\bibitem[{{Lopez} {et~al.}(2012){Lopez}, {Fortney}, \&
  {Miller}}]{LopezEt2012ApJ}
{Lopez}, E.~D., {Fortney}, J.~J., \& {Miller}, N. 2012, \apj, 761, 59

\bibitem[{{Luger} {et~al.}(2015){Luger}, {Barnes}, {Lopez}, {Fortney},
  {Jackson}, \& {Meadows}}]{LugerEt2015AsBio}
{Luger}, R., {Barnes}, R., {Lopez}, E., {et~al.} 2015, Astrobiology, 15, 57

\bibitem[{{MacLeod} {et~al.}(2012){MacLeod}, {Guillochon}, \&
  {Ramirez-Ruiz}}]{MacLeodEt2012ApJ}
{MacLeod}, M., {Guillochon}, J., \& {Ramirez-Ruiz}, E. 2012, \apj, 757, 134

\bibitem[{{Mullally} {et~al.}(2015){Mullally}, {Coughlin}, {Thompson}, {Rowe},
  {Burke}, {Latham}, {Batalha}, {Bryson}, {Christiansen}, {Henze}, {Ofir},
  {Quarles}, {Shporer}, {Van Eylen}, {Van Laerhoven}, {Shah}, {Wolfgang},
  {Chaplin}, {Xie}, {Akeson}, {Argabright}, {Bachtell}, {Barclay}, {Borucki},
  {Caldwell}, {Campbell}, {Catanzarite}, {Cochran}, {Duren}, {Fleming},
  {Fraquelli}, {Girouard}, {Haas}, {He{\l}miniak}, {Howell}, {Huber}, {Larson},
  {Gautier}, {Jenkins}, {Li}, {Lissauer}, {McArthur}, {Miller}, {Morris},
  {Patil-Sabale}, {Plavchan}, {Putnam}, {Quintana}, {Ramirez}, {Silva Aguirre},
  {Seader}, {Smith}, {Steffen}, {Stewart}, {Stober}, {Still}, {Tenenbaum},
  {Troeltzsch}, {Twicken}, \& {Zamudio}}]{MullallyEt2015ApJS}
{Mullally}, F., {Coughlin}, J.~L., {Thompson}, S.~E., {et~al.} 2015, \apjs,
  217, 31

\bibitem[{{Murray-Clay} {et~al.}(2009){Murray-Clay}, {Chiang}, \&
  {Murray}}]{MurrayClayEt2009ApJ}
{Murray-Clay}, R.~A., {Chiang}, E.~I., \& {Murray}, N. 2009, \apj, 693, 23

\bibitem[{{Murray-Clay} \& {Loeb}(2012)}]{Murray-Clay+Loeb2012}
{Murray-Clay}, R.~A., \& {Loeb}, A. 2012, Nature Communications, 3, 1049

\bibitem[{{Nicastro} {et~al.}(2016){Nicastro}, {Senatore}, {Krongold},
  {Mathur}, \& {Elvis}}]{NicastroEt2016ApJ}
{Nicastro}, F., {Senatore}, F., {Krongold}, Y., {Mathur}, S., \& {Elvis}, M.
  2016, \apjl, 828, L12

\bibitem[{{Owen} \& {Jackson}(2012)}]{Owen&Jackson2012ApJ}
{Owen}, J.~E., \& {Jackson}, A.~P. 2012, \mnras, 425, 2931

\bibitem[{{Owen} \& {Wu}(2013)}]{Owen&Wu2013ApJ}
{Owen}, J.~E., \& {Wu}, Y. 2013, \apj, 775, 105

\bibitem[{{Owen} \& {Wu}(2017)}]{Owen+Wu2017ApJ}
---. 2017, \apj, 847, 29

\bibitem[{{Paxton} {et~al.}(2011){Paxton}, {Bildsten}, {Dotter}, {Herwig},
  {Lesaffre}, \& {Timmes}}]{PaxtonEt2011ApJS}
{Paxton}, B., {Bildsten}, L., {Dotter}, A., {et~al.} 2011, \apjs, 192, 3

\bibitem[{{Paxton} {et~al.}(2013){Paxton}, {Cantiello}, {Arras}, {Bildsten},
  {Brown}, {Dotter}, {Mankovich}, {Montgomery}, {Stello}, {Timmes}, \&
  {Townsend}}]{PaxtonEt2013ApJS}
{Paxton}, B., {Cantiello}, M., {Arras}, P., {et~al.} 2013, \apjs, 208, 4

\bibitem[{Pounds {et~al.}(2003)Pounds, Reeves, King, Page, O'Brien, \&
  Turner}]{pounds2003high}
Pounds, K.~A., Reeves, J., King, A.~R., {et~al.} 2003, Monthly Notices of the
  Royal Astronomical Society, 345, 705

\bibitem[{{Ribas} {et~al.}(2005){Ribas}, {Guinan}, {G{\"u}del}, \&
  {Audard}}]{RibasEt2005ApJ}
{Ribas}, I., {Guinan}, E.~F., {G{\"u}del}, M., \& {Audard}, M. 2005, \apj, 622,
  680

\bibitem[{{Rogers}(2015)}]{Rogers2015ApJ}
{Rogers}, L.~A. 2015, \apj, 801, 41

\bibitem[{{Rogers} {et~al.}(2011){Rogers}, {Bodenheimer}, {Lissauer}, \&
  {Seager}}]{RogersEt2011ApJ}
{Rogers}, L.~A., {Bodenheimer}, P., {Lissauer}, J.~J., \& {Seager}, S. 2011,
  \apj, 738, 59

\bibitem[{{Saumon} {et~al.}(1995){Saumon}, {Chabrier}, \& {van
  Horn}}]{SaumonEt1995ApJS}
{Saumon}, D., {Chabrier}, G., \& {van Horn}, H.~M. 1995, \apjs, 99, 713

\bibitem[{{Sch{\"o}del} {et~al.}(2017){Sch{\"o}del}, {Gallego-Cano}, {Dong},
  {Nogueras-Lara}, {Gallego-Calvente}, {Amaro-Seoane}, \&
  {Baumgardt}}]{SchodelEt2017arXiv}
{Sch{\"o}del}, R., {Gallego-Cano}, E., {Dong}, H., {et~al.} 2017, ArXiv
  e-prints, arXiv:1701.03817

\bibitem[{Sigurdsson(1992)}]{Sigurdsson1992ApJ}
Sigurdsson, S. 1992, The Astrophysical journal, 399, L95

\bibitem[{{Spergel} {et~al.}(2015){Spergel}, {Gehrels}, {Baltay}, {Bennett},
  {Breckinridge}, {Donahue}, {Dressler}, {Gaudi}, {Greene}, {Guyon}, {Hirata},
  {Kalirai}, {Kasdin}, {Macintosh}, {Moos}, {Perlmutter}, {Postman},
  {Rauscher}, {Rhodes}, {Wang}, {Weinberg}, {Benford}, {Hudson}, {Jeong},
  {Mellier}, {Traub}, {Yamada}, {Capak}, {Colbert}, {Masters}, {Penny},
  {Savransky}, {Stern}, {Zimmerman}, {Barry}, {Bartusek}, {Carpenter}, {Cheng},
  {Content}, {Dekens}, {Demers}, {Grady}, {Jackson}, {Kuan}, {Kruk}, {Melton},
  {Nemati}, {Parvin}, {Poberezhskiy}, {Peddie}, {Ruffa}, {Wallace}, {Whipple},
  {Wollack}, \& {Zhao}}]{SpergelEt2015arXiv}
{Spergel}, D., {Gehrels}, N., {Baltay}, C., {et~al.} 2015, ArXiv e-prints,
  arXiv:1503.03757

\bibitem[{{Thomas} {et~al.}(2016){Thomas}, {Groves}, {Sutherland}, {Dopita},
  {Kewley}, \& {Jin}}]{ThomasEt2016ApJ}
{Thomas}, A.~D., {Groves}, B.~A., {Sutherland}, R.~S., {et~al.} 2016, \apj,
  833, 266

\bibitem[{{Valencia} {et~al.}(2010){Valencia}, {Ikoma}, {Guillot}, \&
  {Nettelmann}}]{ValenciaEt2010A&A}
{Valencia}, D., {Ikoma}, M., {Guillot}, T., \& {Nettelmann}, N. 2010, \aap,
  516, A20

\bibitem[{{Van Eylen} {et~al.}(2017){Van Eylen}, {Agentoft}, {Lundkvist},
  {Kjeldsen}, {Owen}, {Fulton}, {Petigura}, \& {Snellen}}]{VanEylenEt2017}
{Van Eylen}, V., {Agentoft}, C., {Lundkvist}, M.~S., {et~al.} 2017, ArXiv
  e-prints, arXiv:1710.05398

\bibitem[{{Wyithe} \& {Loeb}(2003)}]{Wyithe2003ApJ}
{Wyithe}, J.~S.~B., \& {Loeb}, A. 2003, \apj, 595, 614

\bibitem[{{Yang} {et~al.}(2014){Yang}, {Bou{\'e}}, {Fabrycky}, \&
  {Abbot}}]{YangEt2014ApJL}
{Yang}, J., {Bou{\'e}}, G., {Fabrycky}, D.~C., \& {Abbot}, D.~S. 2014, \apjl,
  787, L2

\bibitem[{{Zeng} {et~al.}(2017){Zeng}, {Jacobsen}, \&
  {Sasselov}}]{ZengEt2017arXiv}
{Zeng}, L., {Jacobsen}, S.~B., \& {Sasselov}, D.~D. 2017, Research Notes of the
  American Astronomical Society, 1, 32

\end{thebibliography}

\end{document}